\documentclass[11pt]{article}
\usepackage{moriond,epsfig}

\def\be{\begin{equation}}
\def\ee{\end{equation}}
\def\bea{\begin{eqnarray}}
\def\eea{\end{eqnarray}}

\begin{document}
\vspace*{4cm}
\title{Searches for the Standard Model Higgs Boson at the Tevatron}

\author{ Tommaso Dorigo }

\address{Dipartimento di Fisica ``G.Galilei'', Via Marzolo 8, 35131 Padova, Italy}

\maketitle\abstracts{
The CDF and D\O\ experiments at the Tevatron have searched for the Standard Model
Higgs boson in data collected between 2001 and 2004. Upper limits have been
placed on the production cross section times branching  ratio to $b\bar{b}$ pairs
or $W^+ W^-$ pairs as a function of the Higgs boson mass. Projections indicate
that the Tevatron experiments have a chance of discovering a $M_H=115$ GeV Higgs with
the total dataset foreseen by 2009, or excluding it at 95\% C.L. up to a mass of 135 GeV. }

\section{Introduction}

The excellent agreement between experimental measurements of electroweak observables
and Standard Model (SM) predictions constitute a strong motivation to search for 
the Higgs boson at the Tevatron~\cite{gunion}. The latest fits~\cite{lepewwg}, 
which indicate as $M_H=126^{+73}_{-48}$ GeV the most likely
value for the Higgs mass, together with the direct lower limit of $M_H>114$ GeV 
obtained by the LEP experiments~\cite{leplimit}, allow CDF and D\O\ to hope for a 
significant measurement before the Large Hadron Collider at CERN starts producing 
proton-proton collisions at a center-of-mass energy of 14 TeV. 

\begin{figure}[h!]
\epsfig{file=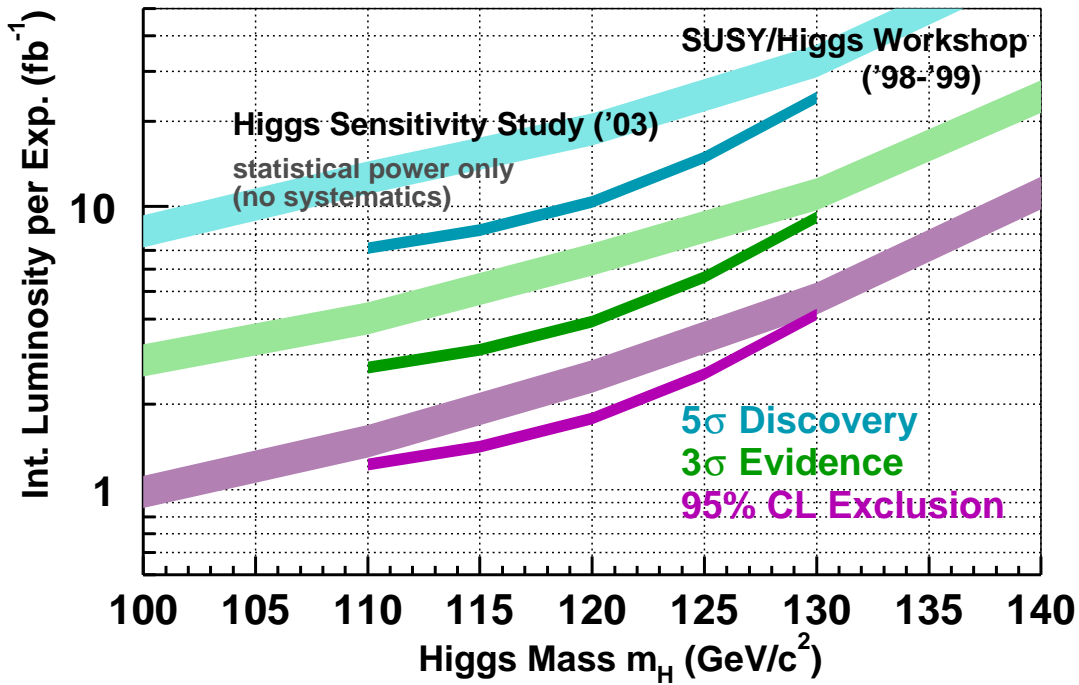,bb=0 0 570 360,width=15cm,clip=}
\caption{ \em The three bands show the integrated luminosity per experiment needed for 
a $95\%$ exclusion (purple), a $3 \sigma$ evidence (green), or a $5 \sigma$ discovery (blue) of the Higgs
boson as a function of the particle mass. The 2003 study, which only considers the low mass regime
($M_H < 135$ GeV), shows slightly reduced luminosity thresholds for a given mass, 
but does not include systematic uncertainties.
\label{f:hswg}}
\end{figure}

In 2003 a joint committee of CDF and D\O\ members carried out 
a reassessment of the Tevatron reach in the search for the
Higgs boson~\cite{hswgrep}. By using a more
realistic model of the two detectors than the simplified one used in a study 
performed in 1998~\cite{hwg}, together with real data collected by the experiments, the committee 
determined that the early claims of sensitivity were not off the mark. In the meantime, the Tevatron
has continued to improve its performance, recently surpassing the peak luminosity of 
$1.2\times 10^{32} cm^{-2} s^{-1}$. The chances of delivering an integrated statistics of
$8 fb^{-1}$ by the end of 2009 appear now sizeable. It seems therefore possible
to expect that before the CMS and ATLAS collaborations start analyzing their first collisions, 
CDF and D\O\ may either discover a 115 GeV Higgs boson, or exclude it up to 135 GeV 
(see Fig.~\ref{f:hswg}).

\noindent
The search for the Higgs boson at the Tevatron is carried out by looking for two different
final states depending on the particle mass. 
For masses below 135 GeV, the dominant decay is $H \to b\bar{b}$,
and the search channels always include an accompanying $W$ or $Z$ boson signature; 
the latest results from CDF and D\O\ in these final states are given in Sec.~\ref{s:light}. 
For masses above that threshold, the $H \to W^+W^-$ decay is the most promising one, and both 
direct production and production in association with an additional electroweak boson are considered; 
results are discussed in Sec.~\ref{s:heavy}. 

\section {Tools }

The smallness of the Higgs boson signal at the Tevatron, when compared with competing backgrounds,
implies that for a successful search many tools have to be crafted to the highest standards: 
a performant lepton identification, efficient $b$-quark tagging, and precise jet energy measurement.

\begin{figure}[h!]
\centerline{\epsfig{file=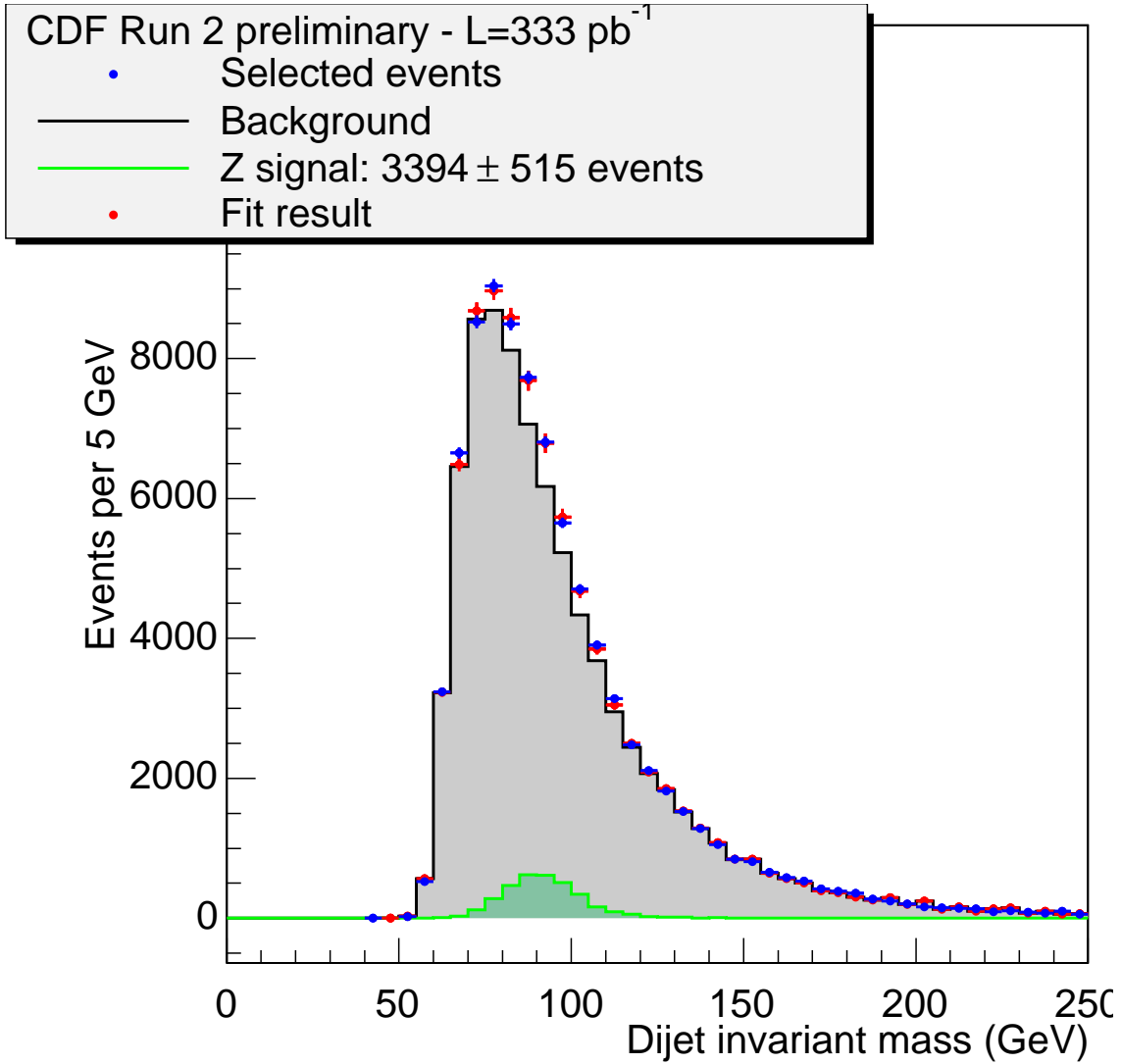,bb=0 0 568 547, width=15cm,clip=}}
\caption{ The signal of $Z \to b\bar{b}$ decays extracted from 333 $pb^{-1}$ of CDF Run 2 data
is visible as an excess of events in the dijet mass region around 90 GeV. The green histogram
shows the fitted $Z$ contribution.
\label{f:zbbsignal}}
\end{figure}

Both CDF~\cite{cdfleptons} and D\O\ ~\cite{d0leptons} excel in detecting high-$P_T$ electrons and muons generated from $W$ or $Z$ boson
decay. Their signal constitutes the trigger for all datasets used in the analyses described in
this paper.

When searching for light Higgs boson decays, the identification and precise measurement of 
$b$-quark-originated jets is crucial. To successfully tag $b$-jets both detectors are endowed with dedicated 
silicon detectors which can measure the impact parameter of charged tracks with a resolution 
of few tens of microns, thereby allowing to determine the decay point of $B$ hadrons. 
Other methods rely on an estimate of the global probability that all tracks in a jet come from 
the primary interaction vertex, 
or on the identification of electrons or muons from the semileptonic decay of $B$ particles. 
A lot of work has gone into perfecting these algorithms,
and both collaborations claim a 45\% to 50\% efficiency to identify secondary vertices in central 
$b$-jets, with fake rates of less than a percent.

The goal for the precise measurement of the energy of $b$-jets is set by 
the Higgs Sensitivity Working Group. They claimed that by using advanced 
algorithms a relative resolution $\sigma_{M_{jj}}/M_{jj} \sim 10\%$
was attainable on the dijet invariant mass pair of $b$-quark jets from $H \to b\bar{b}$ 
decay~\cite{hswgrep}. 
This is still to be demonstrated on the data. The most stringent
tests foresee using the Z boson decay to $b\bar{b}$ pairs, both to check the 
absence of systematics in the determination of the $b$-jet energy scale and to verify the dijet mass
resolution. 

CDF has recently identified a first signal of $Z \to b\bar{b}$ decays from a dedicated dataset,
thanks to a online trigger designed to select low-$E_T$ jets 
containing charged tracks with large impact parameter identified by the silicon vertex 
tracker~\cite{zbbsvt}. Fig.~\ref{f:zbbsignal} shows the
$Z$ signal, whose size and shape are in good agreement with expectations. 
Besides providing a precise measurement of the $b$-jet energy scale, 
the data will be used in studies aimed at improving the dijet mass resolution, to increase the 
chances of a Higgs discovery in the $b\bar{b}$ final state.

\section {Light Higgs Boson Searches \label{s:light}}

Both CDF and D\O\ have analyzed their Run 2 datasets in search for associated production
of a $W$ boson and a pair of $b$-quark jets from $H\to b\bar{b}$ decay, using the dijet mass
distribution to extract limits to the 
cross section times branching ratio of the process.

\begin{figure}[h!]
\epsfig{file=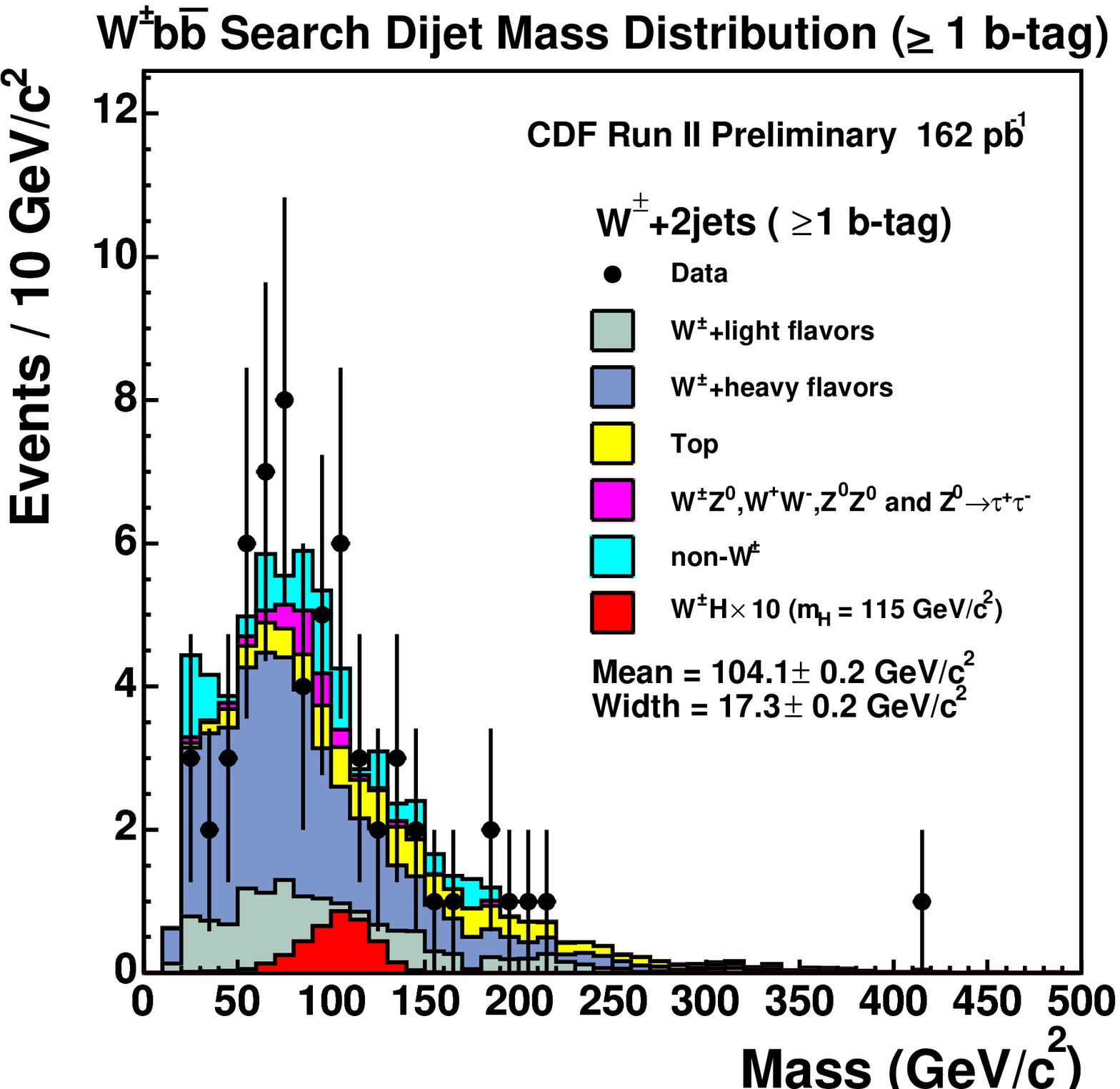, bb=-15 -15 543 515, width=7.5cm, clip=}
\epsfig{file=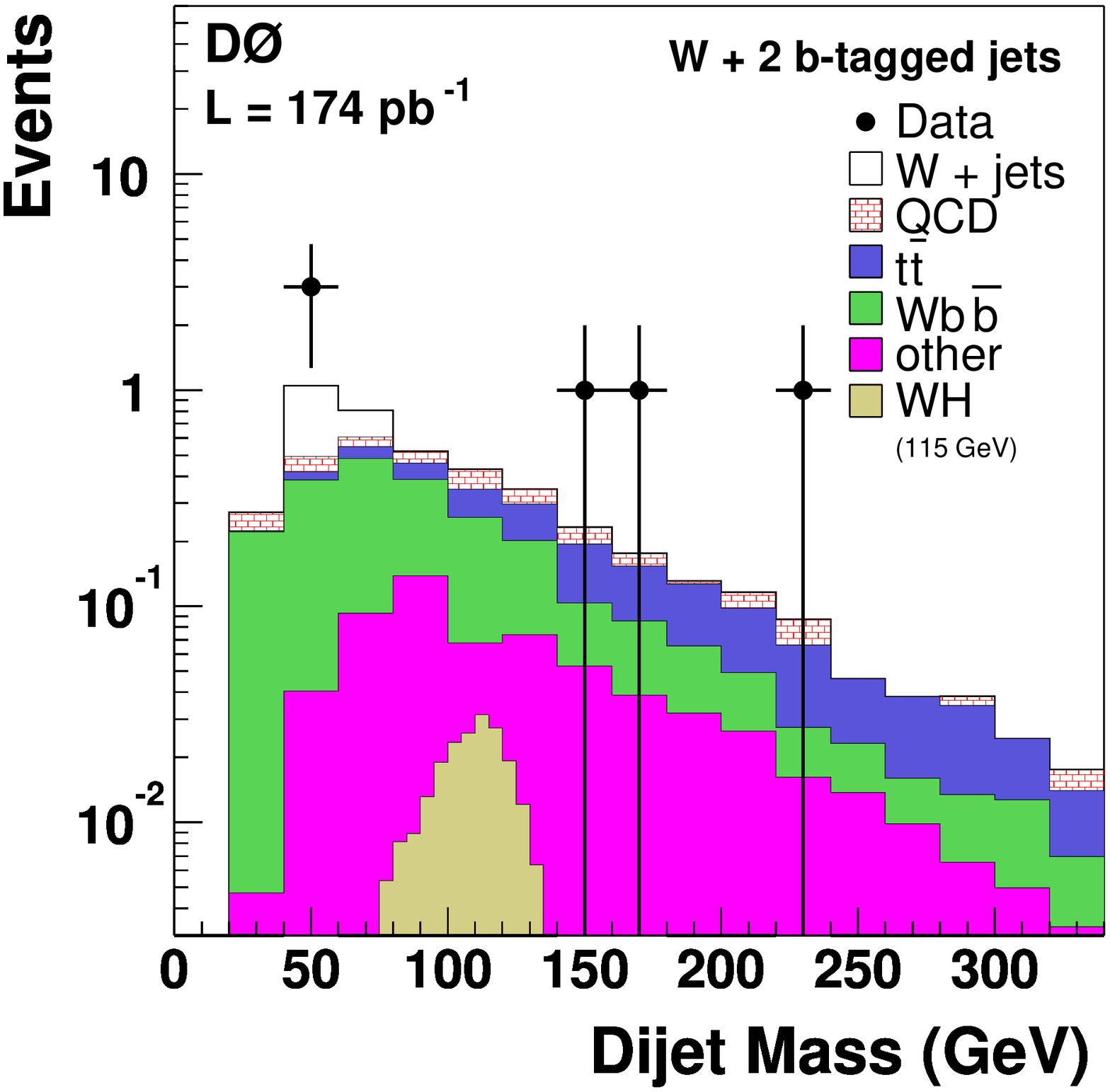, bb=0 5 596 596, width=7.5cm, clip=}
\caption{\em The dijet mass distribution of 62 b-tagged W +2 jet candidates (black points) is
understood as a sum of several contributing SM backgrounds (left); a $H \to b\bar{b}$ signal 10 times
larger than expected is overlaid. Right: 
the dijet mass of double b-tagged W+ 2 jet candidates found by D\O\ ; the expected $H \to b\bar{b}$
signal is overlaid.
\label{f:cdffit_2j}}
\end{figure}

\noindent
In $162 pb^{-1}$ of collider data CDF finds 62 events with a clean $W \to l\nu$ ($l=e, \mu$)
candidate and two jets, one of which tagged by secondary vertex identification; 
estimated SM backgrounds amount to $66.5\pm 9.0$ events (Fig.~\ref{f:cdffit_2j}, left). 
From a fit of the dijet
mass distribution of these events a $95\%$ C.L. limit of about $5 pb$ is obtained, with little 
dependence on the Higgs boson mass (see Fig.~\ref{f:limitsummary}).

D\O\ similarly finds 76 events in $174 pb^{-1}$ of $p\bar{p}$ collisions yielding a $W \to e\nu$ 
signal and two jets, at least one of which with a $b$-tag. Estimated backgrounds amount to 
$72.6 \pm 20.0$ events.
The subset of 6 events with both jets $b$-tagged is then divided in search windows in the dijet 
mass distribution, to extract cross section limits to Higgs boson production 
(Fig.~\ref{f:cdffit_2j}, right). The result is 
a $95\%$ C.L. limit of $\sigma_{WH}\times B(H \to b\bar{b}) <9$ to $12 pb$ for $M_H=115$ to 135 GeV
(see Fig.~\ref{f:limitsummary}).

\section {High Mass Searches \label{s:heavy}}

For $M_H>135$ GeV the $H \to WW^{(*)}$ decay mode becomes dominant. When both $W$ bosons decay to 
an electron-neutrino or muon-neutrino pair the final state is very clean, with residual backgrounds
mostly due to Drell-Yan production of lepton pairs. To discriminate direct production
of a Higgs boson from non-resonant $WW$ production it is useful to study the azimuthal angle 
$\Delta \Phi_{ll}$ between
the two charged leptons, since the zero spin of the Higgs boson and helicity conservation conspire
to produce leptons in the same direction in the transverse plane. 

\begin{figure}[h!]
\epsfig{file=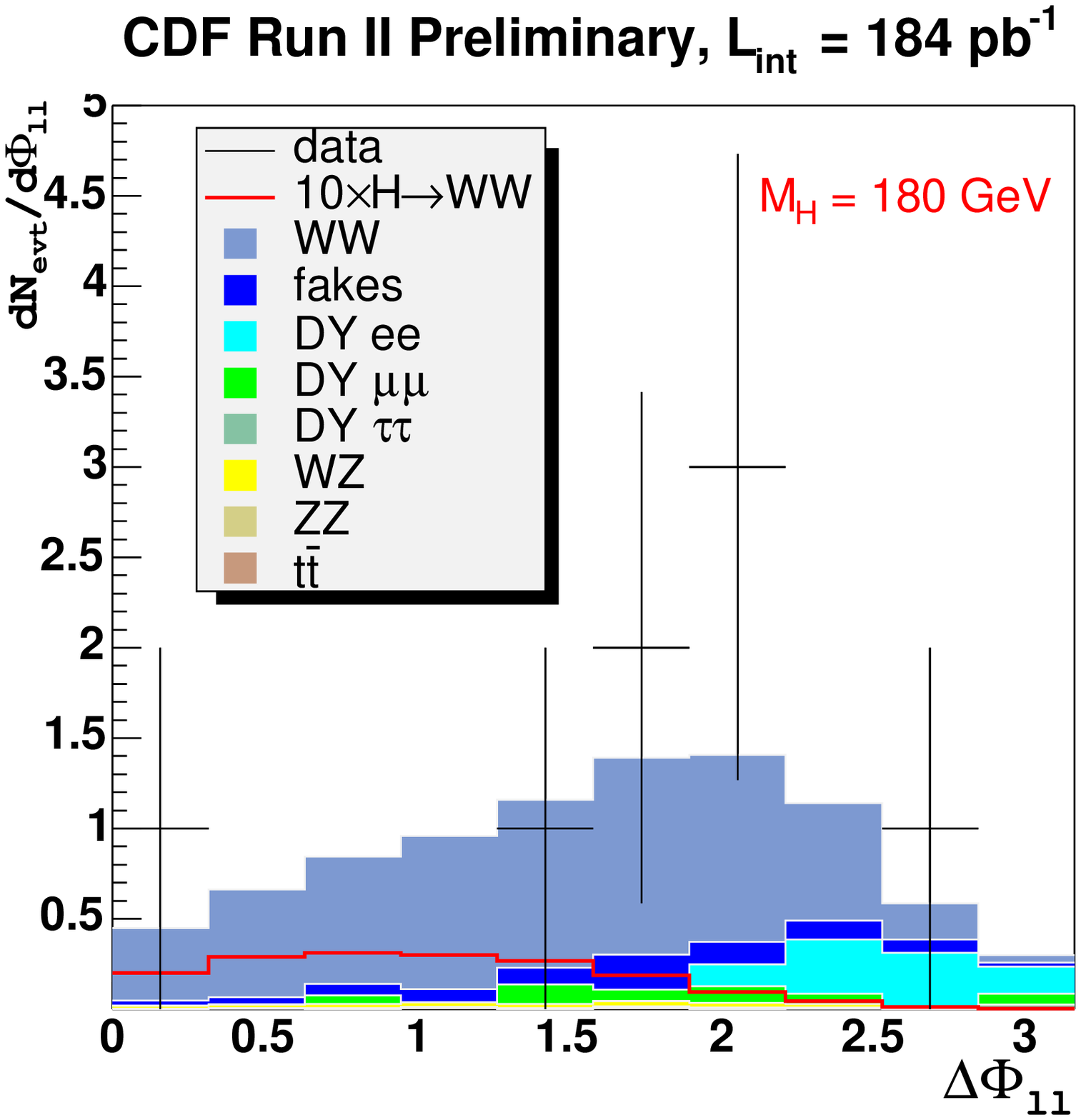, bb=36 22 519 514, width=7.5cm, clip=}
\epsfig{file=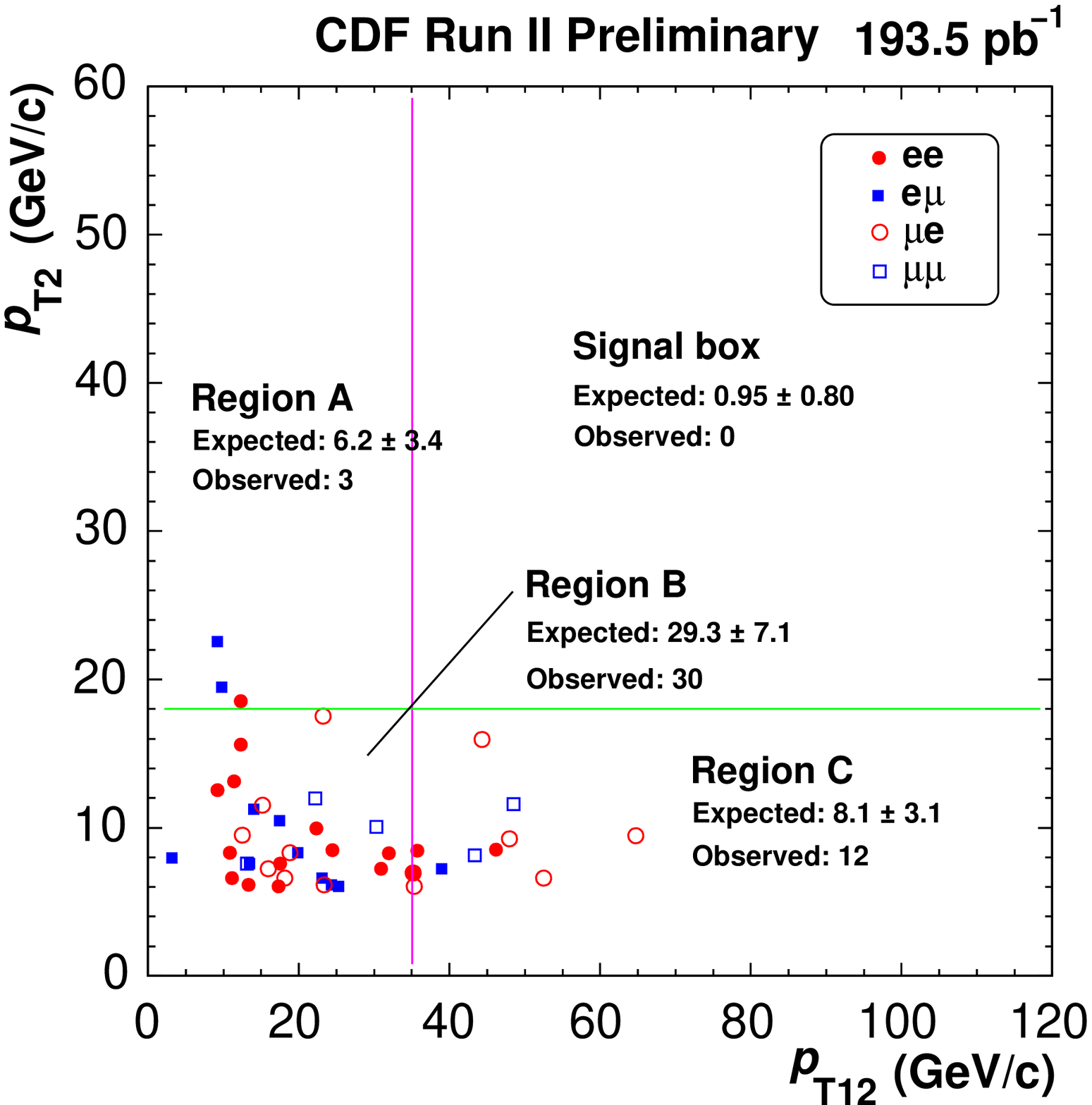, bb=0 0 526 533, width=7.5cm, clip=}
\caption{\em Azimuthal angle between charged leptons in the CDF $H\to WW$ analysis (left).  
Distribution of the vector sum of lepton transverse momenta ($P_{T12}$) versus subleading
lepton transverse momentum ($P_{T2}$) for same-sign dilepton candidates found by CDF (right). 
\label{f:cdf_dphill}}
\end{figure}

CDF selects $W$ pairs by searching for lepton pairs of opposite charge, $P_T>20$ GeV, and then 
applying a missing $E_T$ cut at 25 GeV and a tight jet veto. A small dilepton mass is then required
($M_{ll}<55~(80)$ GeV for $M_H=140~(180)$ GeV). After this selection, 8 events are observed
in $184 pb^{-1}$ of data, where $8.9\pm1.0$ are expected from non-Higgs SM sources. A likelihood fit
to the $\Delta \Phi_{ll}$ distribution is performed to extract a $95\%$ C.L. limit on the
cross section as a function of $M_H$ (see Fig.~\ref{f:cdf_dphill}, left). 
The result is $\sigma_H \times B(H \to WW)<5.6 pb$ for $M_H=160$ GeV.

\noindent
D\O\ similarly searches for $WW$ pairs in $177 pb^{-1}$ of Run 2 data. They require the azimuthal
angle between the charged leptons to be smaller than 2.0 radians, and find 9 $WW$ candidates, when
$11.2\pm3.2$ are expected from non-Higgs SM sources. 
They thus obtain $\sigma_H \times B(H\to WW)<5.7 pb$ at $95\%$ C.L. for $M_H= 160$ GeV.

\begin{figure}[h!]
\centerline{\epsfig{file=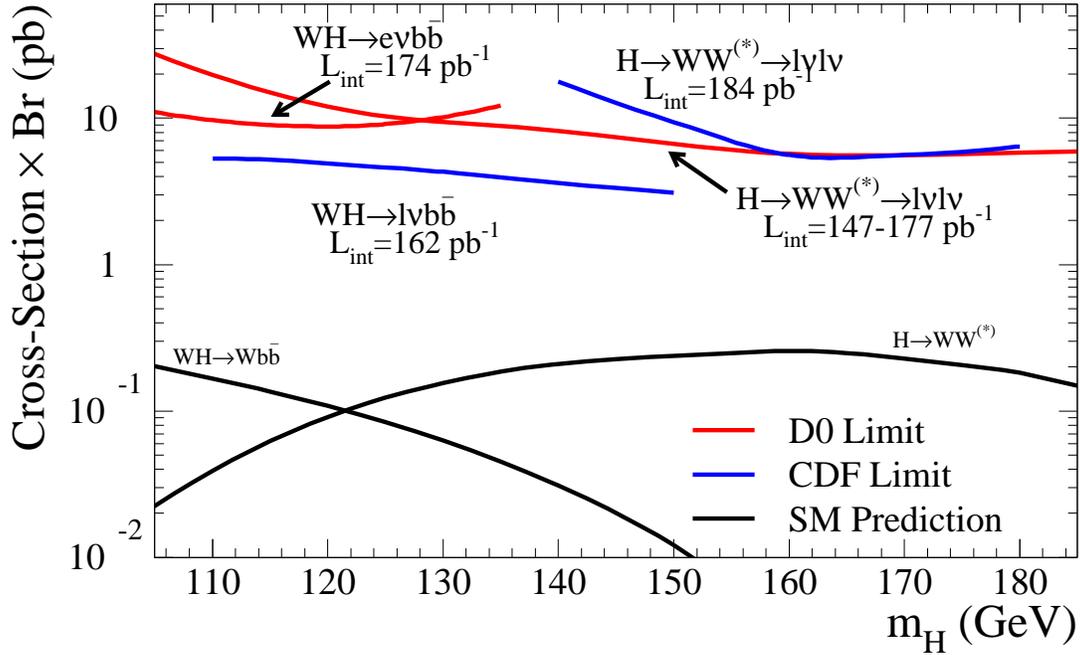,bb=10 35 530 380, width=15cm, clip=}}
\caption{\em Summary of $95\%$ C.L. limits obtained by CDF and D\O\ from Run 2 searches of the Higgs
boson. The expected SM cross section for the searched processes is in black.
\label{f:limitsummary} }
\end{figure}

CDF also searches $193.5 pb^{-1}$ of data for the striking signature of associated $WH$ production 
at high $M_H$, when the event may yield {\em three} $W$ bosons. 
The search starts from a dataset of lepton pairs of
same charge, which is understood as a sum of fake lepton backgrounds and SM sources 
(see Fig.~\ref{f:cdf_dphill}, right). 
Optimized cuts are then applied to the leptons transverse momentum and to the vector sum 
of lepton transverse momenta, $P_{t_{12}}>35$ GeV. 
Zero events are observed, with $1.0\pm0.6$ expected from known sources. 
A $95\%$ C.L. cross section times branching ratio limit of $8 pb$ can thus be set for $M_H=160$ GeV
 (see Fig.~\ref{f:limitsummary}).

\section {Conclusions}

The Higgs boson is being hunted at the Tevatron in all advantageous search channels. If the actual
luminosity delivered by the accelerator keeps following expectations, the Tevatron experiments
have a real chance of observing a $M_H=115$ GeV Higgs boson before the end of 2009, or exclude its
existence if $M_H < 135$ GeV.

\section*{Acknowledgments}
The author wishes to thank Demie Cheng for her editorial advice.

\end{document}